The Final Manuscript to Oxford Science Encyclopedia:

# The formation of the Martian moons

Rosenblatt P., Hyodo R., Pignatale F., Trinh A., Charnoz S., Dunseath K.M., Dunseath-Terao M., & Genda H.

## Summary

Almost all the planets of our solar system have moons. Each planetary system has however unique characteristics. The Martian system has not one single big moon like the Earth, not tens of moons of various sizes like for the giant planets, but two small moons: Phobos and Deimos. *How did form such a system?* This question is still being investigated on the basis of the Earth-based and space-borne observations of the Martian moons and of the more modern theories proposed to account for the formation of other moon systems.

The most recent scenario of formation of the Martian moons relies on a giant impact occurring at early Mars history and having also formed the so-called hemispheric crustal dichotomy. This scenario accounts for the current orbits of both moons unlike the scenario of capture of small size asteroids. It also predicts a composition of disk material as a mixture of Mars and impactor materials that is in agreement with remote sensing observations of both moon surfaces, which suggests a composition different from Mars. The composition of the Martian moons is however unclear, given the ambiguity on the interpretation of the remote sensing observations.

The study of the formation of the Martian moon system has improved our understanding of moon formation of terrestrial planets: The giant collision scenario can have various outcomes and not only a big moon as for the Earth. This scenario finds a natural place in our current vision of the early solar system when conditions were favorable for giant collisions to occur. The next step in exploration of Martian moon is a sample return mission to test the giant collision scenario for their origin, and to provide tests of models of early solar system dynamics since Mars may retain material exchanged between the inner and outer solar system.

## Introductory paragraphs

The origin of the natural satellites or moons of the solar system is as challenging to unravel as the formation of the planets. Before the start of the space probe exploration era, this topic of planetary science was restricted to telescopic observations, which limited the possibility of testing different formation scenarios. This era has considerably boosted this topic of research, particularly after the Apollo missions returned samples from the Moon's surface to Earth. Observations from subsequent deep space missions such as Viking Orbiter 1 & 2, Voyager 1 & 2, Phobos-2, Galileo, Cassini-Huygens, and the most recent Mars orbiters such as Mars Express as well as from the Hubble space telescope have served to intensify research in this area.

Each moon system has its own specificities, with different origins and histories. It is widely accepted that the Earth's Moon formed after a giant collision between the proto-Earth and a body similar in size to Mars. The Galilean moons of Jupiter, on the other hand, appear to have formed by accretion in a circum-Jovian disk, while smaller, irregularly-shaped satellites were probably captured by the giant planet. The small and medium-sized Saturnian moons may have formed from the rings encircling the planet. Among the terrestrial planets, Mercury and Venus have no moons, the Earth has a single large moon and Mars has two very small satellites. This raises some challenging questions: what processes can lead to moon formation around terrestrial planets, and what parameters determine the possible outcomes, such as the number and size of moons? The answer to such fundamental questions necessarily entails a thorough understanding of the formation of the Martian system, and may have relevance to the possible existence of (exo)moons orbiting exoplanets. The formation of such exomoons is of great importance as they could influence conditions for habitability, or for maintaining life over long periods of time at the surface of Earth-like exoplanets, for example by limiting the variations of the orientation of the planet's rotation axis and thus preventing frequent changes of its climate.

This article summarizes our current knowledge concerning the origin of Phobos and Deimos, acquired from observational data as well as theoretical work. It describes why early observations led to the idea that the two satellites were captured asteroids and why the difficulties in reconciling the current orbits of Phobos and Deimos with those of captured bodies calls for an alternative theory. A detailed description of a giant-impact scenario is then given, in which moons similar to Phobos and Deimos can be formed in orbits similar to those observed today. This scenario also restricts the range of possible composition of the two moons, providing a motivation for future missions that aim for the first time to bring material from the Martian system back to Earth.

### 1- Puzzling origin: Capture vs in-situ formation

#### 1-1  *The Martian moon system*

The two natural satellites of Mars, Phobos and Deimos, were discovered in August 1877 by the American astronomer Asaph Hall[1] (Hall, 1878). They probably went unnoticed for a long time because of their low albedo and their proximity to Mars which makes them difficult to observe with Earth-based telescopes (Pascu et al., 2014). Subsequent telescopic observations determined that their orbits are near-equatorial and near-circular, with Phobos orbiting below the synchronous limit and Deimos orbiting above it (see Table 1). Furthermore, the secular acceleration of Phobos' longitude along its orbit, discovered by Sharpless (1945), indicates that its orbit is slowly decaying, with Phobos losing orbital energy and gradually approaching Mars due to tidal dissipation inside the planet (e.g. Burns, 1992). In contrast, Deimos lies above the synchronous limit and thus is receding from the planet, just as the Moon is receding from Earth. Deimos' secular acceleration has not yet been observed due to its low mass and greater distance from Mars (see Tables 1 & 2). On the basis of the brightness of the two moons, their sizes were first estimated to be a few tens of kilometers (e.g. Pascu et al., 2014). Phobos and Deimos are thus much smaller than the Earth's Moon, the Galilean satellites of

[1] Asaph Hall was greatly supported by his wife Angeline Stickney in his quest of Martian satellites.

Jupiter and the main moons of the other giant planets. Earth-based historical observations have thus highlighted the main challenge raised by the Martian system: how to explain the formation of two small moons in near-equatorial and near-circular orbits around their primary planet?

Table 1: Orbital architecture of the Martian system (Jacobson & Lainey, 2014). The synchronous limit is at 20400 km or around 6 $R_M$, where 1 $R_M$ is Mars' mean radius, equal to 3400 km. This limit is the orbital position where the revolution period around Mars is equal to the spin of the planet (24h39').

|  | Phobos | Deimos |
| --- | --- | --- |
| Semi-major axis | 9375 km (2.76 $R_M$) | 23458 km (7 $R_M$) |
| Eccentricity | 0.01511 | 0.00027 |
| Inclination to the equator | 1.076° | 1.789° |
| Orbital period | 7h 39' 19.47" | 30h 18' 1.36" |
| Secular acceleration along the orbit | 1.273 +/- 0.003 mdeg/year$^2$ | - |

Table 2: Bulk properties of the Martian moons. (1) Willner et al., 2014; (2) Thomas, 1993; (3) Paetzold et al. (2014); (4) Jacobson (2010); (5) using mass from (4) and volume from (1); (6) Rosenblatt, 2011. The radii are from the best fit ellipsoid to the shape of the body (1).

|  | Phobos | Deimos |
| --- | --- | --- |
| Radius (in km) | 13.03 x 11.40 x 9.14 (1) | 7.5 x 6.1 x 5.2 (2) |
| Mass (in $10^{16}$ kg) | 1.066 +/- 0.013 (3) | 0.151 +/- 0.003 (4) |
| Volume (in km$^3$) | 5742 +/- 35 (1) | 1017 +/- 130 (2) |
| Density (in g/cm$^3$) | 1.856 +/- 0.034 (5) | 1.48 +/- 0.22 (6) |

### 1-2   Are Phobos and Deimos small asteroids ?

The investigation of the Martian moon system entered a new era with space probe explorations (Duxbury et al., 2014). The first missions to have made important observations of the two moons were Mariner 9, the Viking 1 and 2 orbiters and Phobos-2. These provided the first resolved images and spectroscopic observations of the surface of Phobos and Deimos (more recent images are shown in figures 1 & 2 and more up-to-date spectra in figure 3), as well as the first determination of their density (mass and volume, see updated values in Table 2). The data suggested that the two small moons resemble asteroids with irregular shape, cratered surface, low albedo (Figures 1 & 2) and low density (Table 2). Furthermore, the reflectance spectra of their surfaces in the near-infrared and visible wavelengths, around 0.4 to 4 microns, show an increasing slope toward infra-red wavelengths (reddening slope, figure 3) that match those of several primitive low-albedo asteroids (Pang et al., 1978; Pollack et al., 1978; Murchie et al., 1991; Murchie & Erard, 1996). The Phobos surface shows areas with different spectral slopes defining the so-called red and blue units while Deimos surface shows only spectra similar to the Phobos red unit (Figure 2 & 3). The reflectance spectra of Phobos have been confirmed by the recent Mars Reconnaissance Orbiter (Fraeman et al., 2012), Mars Express (Fraeman et al., 2012; Witasse et al., 2014) and Rosetta (Pajola et al., 2012) missions as well as by observations obtained with the Hubble space telescope (Rivkin et al., 2002) and with the ground-based Mayall 4-m telescope of the Kitt Peak National Observatory (Fraeman et al., 2014). In addition, Mars Express has confirmed the low density of Phobos by improving estimates of its mass (Paetzold et al., 2014) and volume (Willner et al., 2014). The morphological, physical and spectroscopic similarities between the Martian moons and the primitive asteroids are the main arguments favouring the capture scenario, in which Phobos and Deimos are small low-albedo asteroids captured by the planet (e.g. Pang et al., 1978; Burns, 1992; Pajola et al., 2012; 2013). These asteroids would have formed from the condensation of carbonaceous material in the solar nebula beyond Mars' orbit.

No satisfactory spectral match between Phobos and Deimos and meteoritic material, recognized as the material of primitive asteroids, has however been found so far (e.g. Murchie et al., 1991; Murchie & Erard, 1996, Vernazza et al., 2010). In addition, the reflectance spectra do not clearly display absorption bands (Figure 3) which provide diagnostics for the composition, in terms of carbonaceous, silicate or hydrated material. Space weathering (alteration in space environment over time) on airless bodies may nonetheless suppress any of such possible bands. It is known that this process is responsible for darkening, reddening and suppression of absorption bands (as observed for Phobos and Deimos spectra, figure 3) in reflectance spectra of silicate material, for example on the Moon (Clark et al., 2002; Pieters et al., 2014). It may therefore not be possible to determine the true surface composition of the two moons through spectroscopic observations alone (Rosenblatt 2011). Murchie & Erard (1996) indeed did not rule out that Phobos' spectra could be those of very highly weathered silicate material. It has been proposed more recently that dark silicate (anorthosite) terrains on the Moon can have low albedo similar to those of Phobos and Deimos (Yamamoto et al., 2018). On the other hand, a silicate-rich Phobos would not preclude the possibility of capture, since silicate asteroids can condense from material in the solar nebula closer to Mars' orbit than the region in which carbonaceous asteroids form.

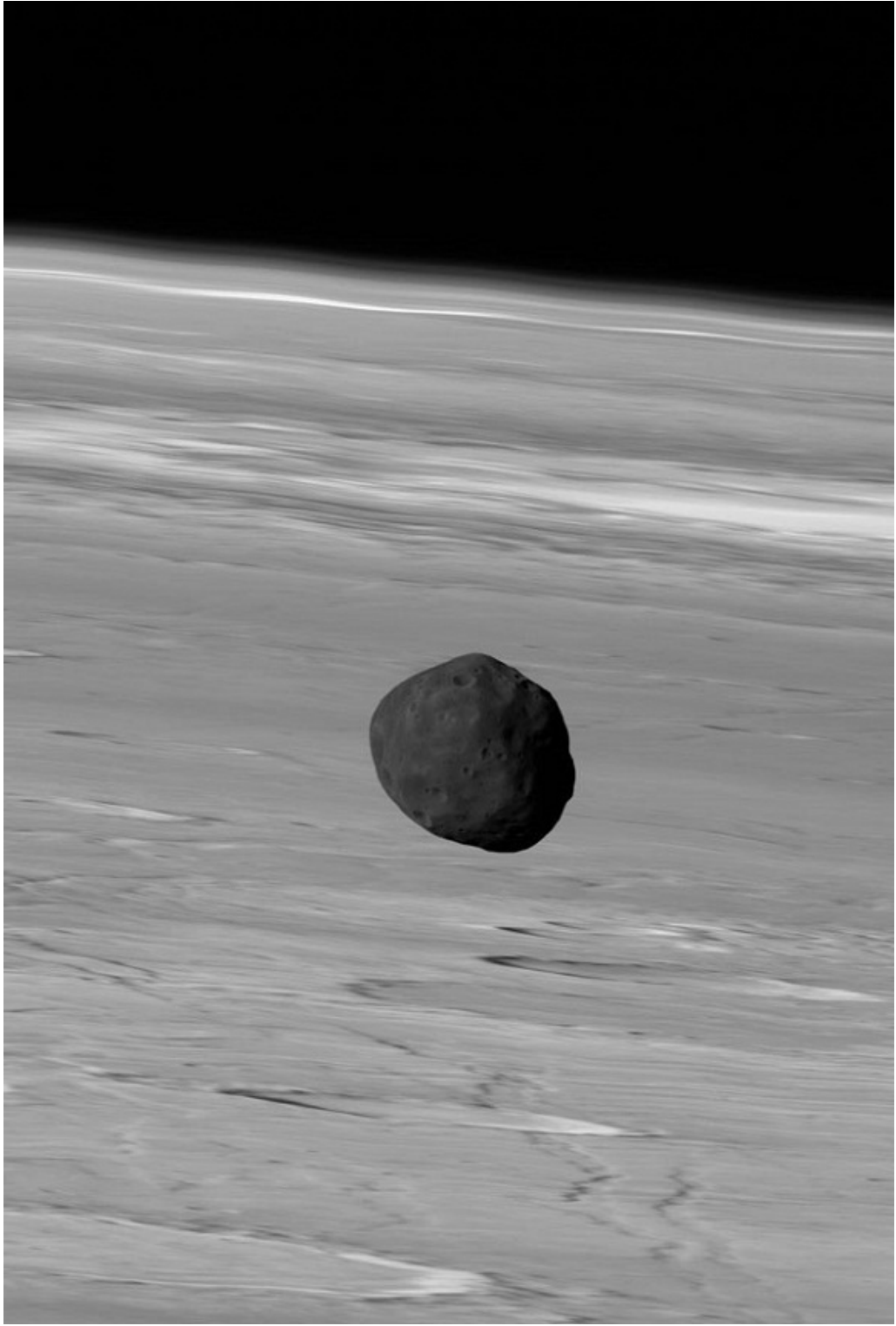

Figure 1: Phobos above Mars as seen by Mars Express, showing the darkness (low albedo) of the moon compared to the brightness of the planet (credit DLR/FU Berlin/ESA).

A tentative, very weak absorption band at around 0.65 microns in the spectra of Phobos has recently been detected (Pajola et al., 2013; Fraeman, et al. 2014; see also figure 3) and could be interpreted as the signature of a carbonaceous composition, although space weathering effects cannot be ruled out (Fraeman et al., 2014). Such effects on carbonaceous material are less well understood than on silicate material, and may give rise to red and even blue spectral slopes observed in reflectance spectra of carbonaceous asteroids (Lantz et al., 2013). The effect of space weathering on reflectance spectra was tentatively simulated for carbonaceous meteorite samples, but the result shared little similarity with Phobos and Deimos (Moroz et al., 2004; Vernazza et al., 2010). Emissivity spectra in the thermal infra-red domain (wavelengths from 5 to 50 microns) of the surface of Phobos were also measured for the first time by Mars Odyssey (Roush and Hogan, 2000) and Mars Express (Giuranna et al., 2011; Witasse et al., 2014). In contrast to the reflectance spectra, these emissivity spectra show clear features more typical of silicate rather than carbonaceous meteorite material (Figure 3, Giuranna et al., 2011). They have however a coarser spatial resolution than the reflectance spectra and are also more affected by grain-size particles of the surface regolith, and to some extent by space weathering (Pieters, 2014).

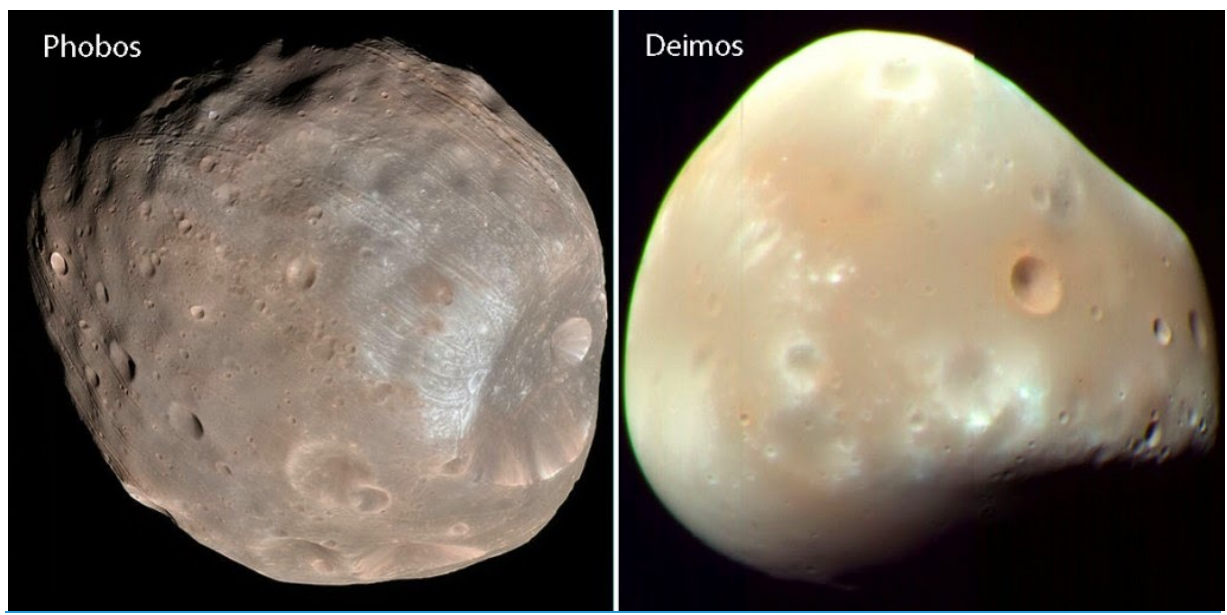

Figure 2: Recent images of the Martian moons from the High Resolution Imaging Science Experiment (HiRISE) onboard the Mars Reconnaissance Orbiter. The individual colour images in near infra-red and blue-green channels have been combined to produce a false-colour representation where patches of high infrared reflectance appear in red while blue-green reflectance patches appear in blue (Thomas et al., 2011). Colour heterogeneities on the surface of the two moons may reveal variations of composition and/or of space weathering effects (Pieters, 2014). The 9 km impact crater Stickney on Phobos appears on the right hand side of the image (Credit NASA/JPL/University of Arizona).

The ambiguity in the composition of the Martian moons derived from remote sensing spectral observations raises an intriguing question: are Phobos and Deimos made of asteroid material (either silicate, carbonaceous or something else) or do they incorporate some Martian material? The latter would suggest that Phobos and Deimos could have been formed in situ around Mars, thus weakening the asteroid capture scenario. In addition, the current near-equatorial and near-circular orbits of the two moons are unlikely to result from a capture (Burns, 1992; Rosenblatt, 2011).

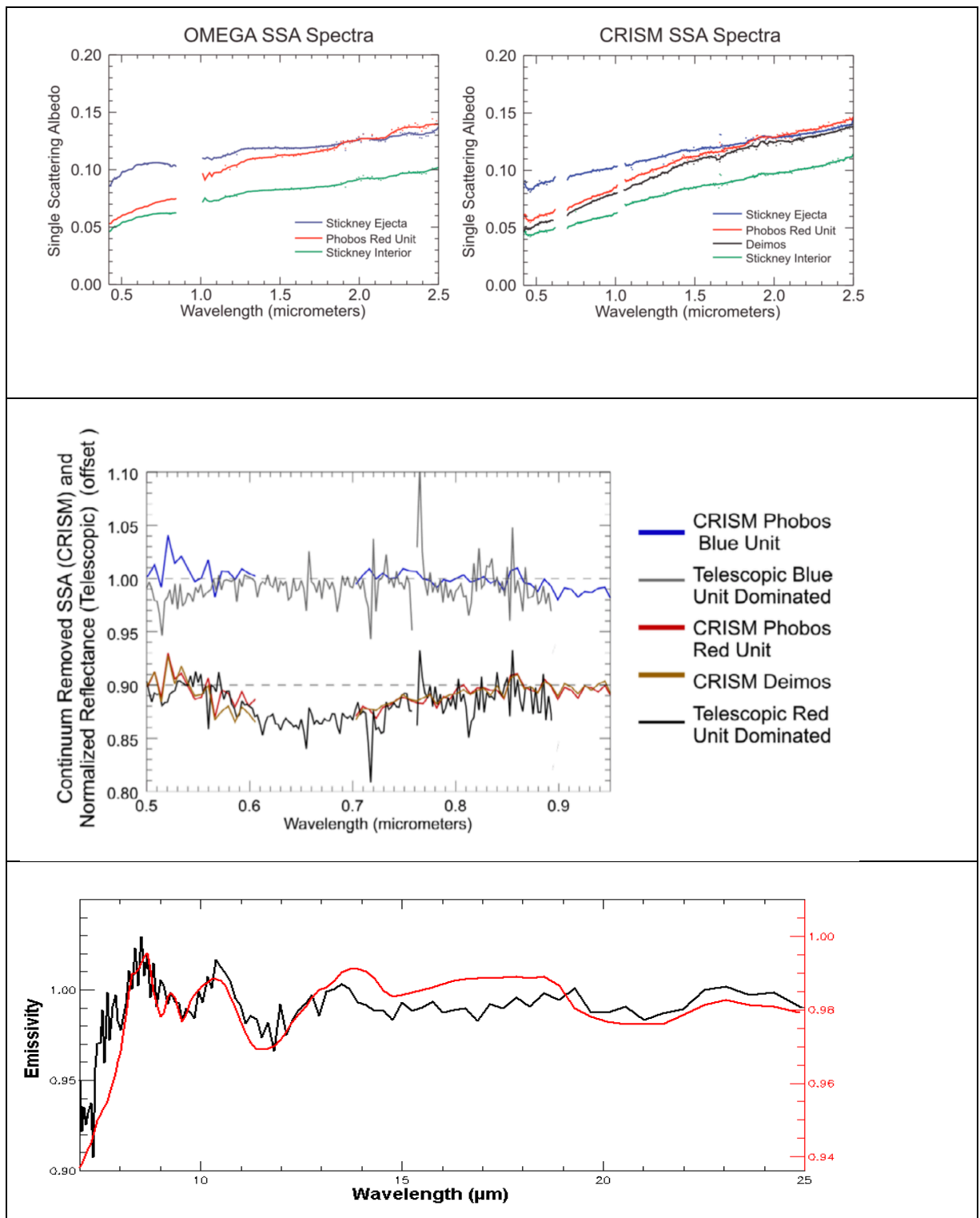

Figure 3: (Top) Most recent Reflectance (Solar Radiation Reflectance) spectra of Phobos and Deimos obtained by the Mars Express' OMEGA and the Mars Reconnaissance Orbiter's CRISM instruments. The Reflectance (Single Scattering Albedo) Spectra of the Stickney Crater area (Blue Unit) is flatter than that of Phobos Red Unit and of Deimos. Adapted from Fraeman et al., 2012. (Medium) CRISM and telescopic observations showing the tentative tiny absorption band at around 0.65 microns for Deimos and for the Red Unit of Phobos, which could be interpreted as the signature of carbonaceous material. Adapted from Fraeman et al. (2014). (Bottom) Emissivity (Thermal Radiation Emission) spectra of Phobos surface (black curve) compared to spectra of silicate material (red curve). Adapted from Giuranna et al. (2011).

### *1.3 Is capture dynamically possible?*

While specific dynamical conditions are required for capture to occur (Pajola et al., 2012), it is not impossible to trap asteroids in a closed orbit around Mars. It can even be facilitated by a three-body capture mechanism, particularly in the early solar system when there was an abundance of planetesimals and other small debris (Hansen, 2018). The orbit of such a captured body is however expected to be significantly elliptical and non-equatorial, as is the case for example for the irregular satellites of Jupiter, in contrast to the actual orbits of Phobos and Deimos (see Table 1). The capture scenario thus requires a mechanism to change the post-capture orbit into the current near-equatorial and near-circular orbits observed today.

One possible mechanism is orbital tidal dissipation (Kaula, 1964). The tides raised by the captured satellite inside Mars and the tides raised by Mars inside the satellite contribute to the dissipation of the satellite orbital energy, and hence modify its orbit. Studies have shown however that this mechanism is not sufficient to change an elliptical and non-equatorial orbit into the almost circular and equatorial orbit of Deimos within the 5 billion years lifetime of the solar system (Szeto, 1983). Since Phobos is larger than Deimos (see Table 2), tidal dissipation has more effect on its orbit: a very elliptical post-capture orbit can be circularized over 5 billion years (e.g. Lambeck, 1979; Cazeneuve et al., 1980; Burns, 1992), assuming a bulk rigidity of rocky material with some degree of micro-porosity (for example, a carbonaceous chondrite material; Lambeck 1979) and a low tidal dissipation factor (Rosenblatt, 2011). Changing the inclination from the ecliptic plane (i.e. the mean orbital plane of the asteroids) to the equatorial plane however requires an even more dissipative material (Mignard, 1981), closer to icy rather than rocky material (Rosenblatt, 2011; Rosenblatt and Pinier, 2014).

A number of possible solutions to this problem have been proposed. One suggestion is that the small body was captured in an equatorial orbit; this however requires that the orbital distance to Mars decreases rapidly below roughly 13 $R_M$ after capture in order to maintain the orbit in the equatorial plane, which seems difficult given how slowly the orbital eccentricity is modified by tidal effects (Burns, 1992). Others authors have proposed that the post-capture orbit was rapidly inclined into the equatorial plane and significantly circularized by drag dissipation in a primitive planetary nebula (Sasaki, 1989). Such a nebula would have been formed around a planet accreting from the solar nebula gas. Drag effect studies (Sasaki, 1989) however have so far been unable to show whether the density profile and survival time of this nebula are consistent with the requirements of the capture scenario.

### *1.4 Alternative scenarios: in-situ formation.*

The difficulty of reconciling the outcome of a capture scenario with the current orbital properties of the Martian moons has motivated the search for alternatives. Most of these assume that the moons have accreted in an equatorial disk of debris containing extra-Martian material in order to explain their possible primitive composition (Rosenblatt, 2011).
This idea of a gravitational aggregate of debris as the bulk structure of the Martian moons is based on their low bulk density (see Table 2), suggesting a large amount of porosity in their interior (Murchie et al., 1991; Andert et al., 2010; Rosenblatt, 2011), as well as on plausible explanations for Phobos' main geomorphological surface features such as the large impact crater Stickney (Bruck Syal et al., 2016) and some of its grooves (Hurford et al., 2016). These surface features would indeed require low rigidity that can be accounted for by porosity (Jaeger et al., 2007; Le Maistre et al., 2013).

Different mechanisms for the formation of the disk have been proposed. One idea is that a body much more massive than Phobos and Deimos might first have been captured by Mars. Its orbit would have rapidly decayed due to tidal forces and the body would have been destroyed when crossing the Roche limit (Singer, 2003). The resulting debris would have formed a ring around Mars below the Roche limit (as modelled later by Black & Mittal, 2015), from which small moons would have emerged. This relatively simple scenario however leads to another impasse: Mars' Roche limit is about three $R_M$, well inside the synchronous limit which is about six $R_M$ (Table 1). It is difficult to find a mechanism by which the moons formed at the Roche limit could have migrated far from Mars and then stayed in orbit for a long time, in particular for Deimos which orbits at a distance of nearly seven $R_M$ (see table 1, Rosenblatt & Charnoz, 2012, see also Section: 'Evolution of the accretion disk').

Another possibility is that the disk resulted from a giant impact, similar to that which led to the formation of the Earth's Moon. This was first proposed by Craddock (1994) but was disregarded by the scientific community until new data about Phobos' interior by Mars Express implying a significant amount of porosity and favoring a formation in an accretion disk (e.g. Andert et al., 2010; Rosenblatt, 2011) revived this scenario. A reapprasial of Phobos' origin from the data of the OMEGA spectrometer onboard Mars Express also favored such a scenario (Bibring, 2010). Craddock's (2011) study suggests that a large body (one quarter to one third the size of Mars) collided with the proto-Mars at least 4 billion years ago, blasting debris into space. Phobos and Deimos may be the last surviving moons that emerged from the resulting accretion disk (Figure 4). The problem still remains, however, of reconciling the migration of moons formed at the Roche limit with the current orbits of Phobos and particularly of Deimos today (Rosenblatt & Charnoz, 2012).

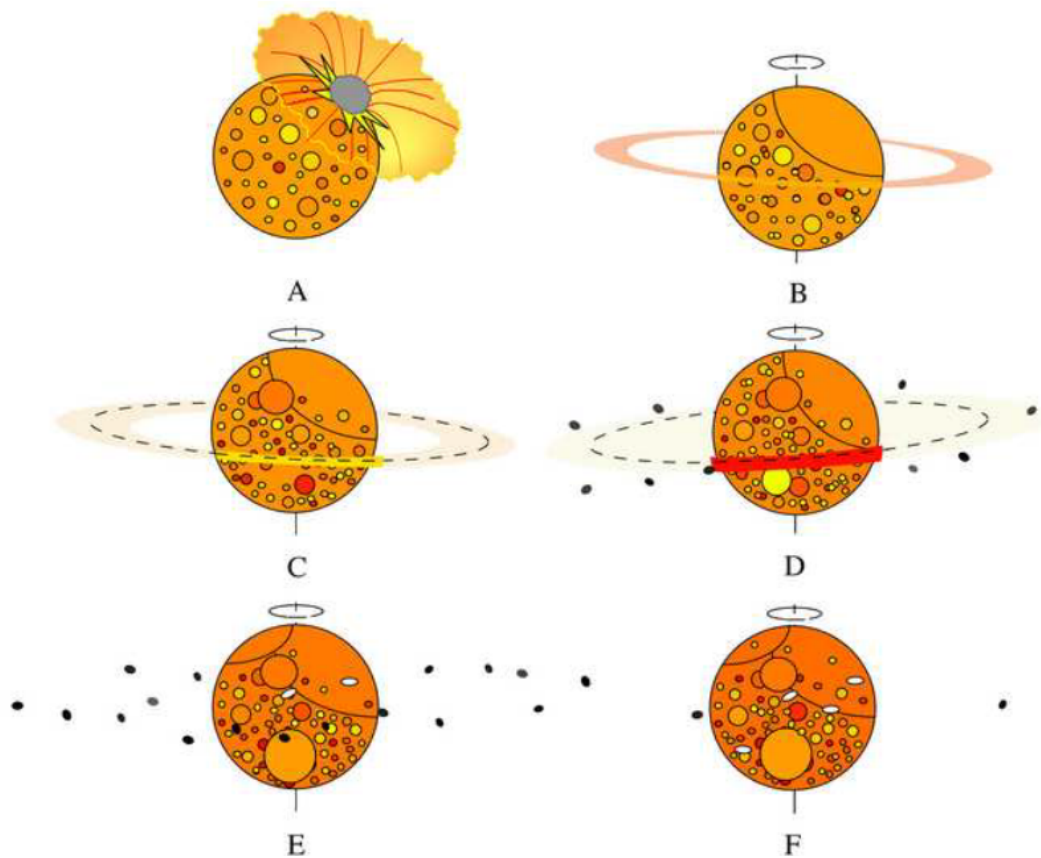

Figure 4: Sketch of the giant impact scenario leading to the formation of Phobos and Deimos (from Craddock, 2011).

## 2- A giant collision scenario

The giant impact scenario leading to the formation of moons can be divided into a number of stages (Figure 4), each of which must be carefully examined in order to build a robust model of the formation process, capable of reproducing the observed orbits and of providing information on the composition of the resulting moons.

### 2-1 A giant impact early in Mars history.

The first question to be answered is whether or not there is any evidence for a giant impact; what effect would such an impact have on the planet and what traces would it leave today?

A giant impact (Figure 5) is thought to be responsible for Mars' current spin rate (Dones and Tremaine, 1993; Craddock, 2011); studies suggest that the mass of the impactor was at least 2% that of Mars, of the order of $10^{22}$ kg. Such an impact would also leave behind a large crater, which would subsequently have been filled to form a relatively flat basin: Borealis, Elysium and Daedalia basins have been identified as possible candidates (Craddock, 2011). An analysis of the elongated crater population on the surface of Mars, supposing that they are all due to impacts of debris from the disk or from the decaying orbit of moons (Schultz and Lutz-Garihan, 1982), suggested that the mass of the disk was of the order of $10^{19}$ kg (Craddock, 2011). However, this elongated crater population may also result from grazing asteroid impacts (Bottke et al., 2000), which would allow for less massive disks.

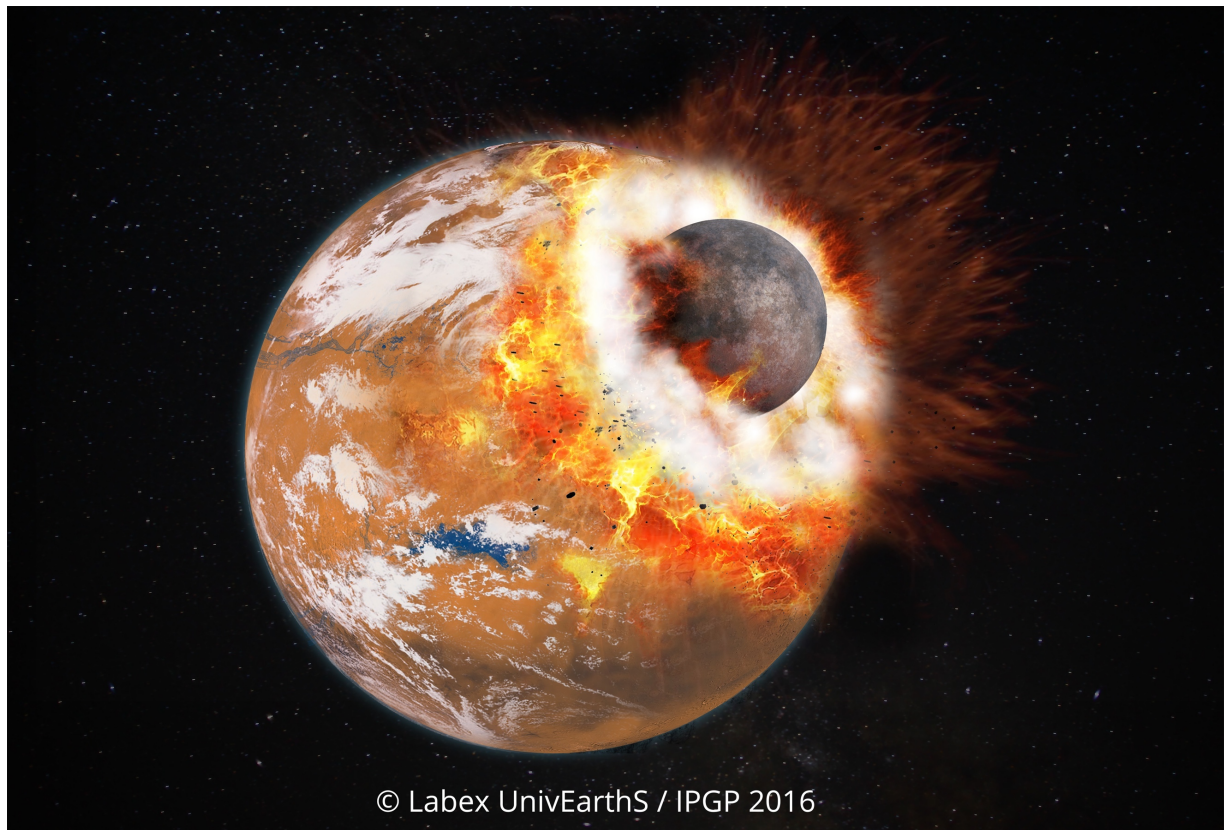

Figure 5: Illustration of a giant collision between the Proto-Mars and a smaller size body (1/3 to 1/4 the size of the impacted planet). Copyright Labex UnivEarthS – USPC – 2016.

On the other hand, the Borealis basin is the largest potential impact basin, measuring 10600 km by 8500 km and covering over 40% of the Martian surface, corresponding to the major hemispheric crustal dichotomy identified on the topography and gravity maps of Mars (Marinova et al., 2008). Numerical models based on the Smoothed Particle Hydrodynamics (SPH) approach conclude that a crater of the size of the Borealis basin can be formed by an object of mass 2.6% that of Mars, moving at about 6 km/s and impacting the surface at 45° (Marinova et al., 2008). These parameters are compatible with the kinds of impact thought to have occurred in the early history of the solar system (Wilhelms and Squyres, 1983) and can account for the spin rate of Mars (Dones and Tremaine, 1993). They have been used in subsequent, chemistry-focused SPH calculations of the nascent accretion disk just after the giant impact (Citron et al., 2015; Hyodo et al., 2017a; Pignatale et al., 2018). Others authors (Canup & Salmon, 2018) however argued for an impactor about ten times less massive with similar impact angle and a slightly higher impact velocity (7 km/s instead of 6 km/s), although still within the range of impactor size/energy estimates to form Borealis. However, these impact conditions cannot fully account for the spin of Mars.

*2-2   Post-impact dynamical evolution of the debris cloud blasted in Mars' orbit: Formation of an accretion disk.*

Immediately after the giant impact, the ejecta that will eventually form Phobos and Deimos have highly elliptical orbits around Mars, with eccentricities between 0.1-0.9, and generally move at velocities different from those of their neighbours (Citron et al., 2015; Hyodo et al. 2017a; Canup and Salmon 2018). At this stage, since the typical temperature is around 2000 K (Figure 6), the ejecta are mostly in the form of molten droplets whose size is determined by the interplay between their differential or shear velocity and the tension on the surface of the droplet. Assuming a surface tension of 0.3 N/m for a silicate melt yields a typical droplet size just after the giant impact of about 1.5 m (Hyodo et al. 2017a). The droplets will quickly solidify since their cooling time, several tens of minutes, is relatively quick compared to their orbital period. As they orbit, the ejecta may collide with each other and undergo further fragmentation, resulting in grains of the order of 100 micron in size (Hyodo et al. 2017a). The eccentricity of the orbits is damped by such collisions and eventually a thin circular disk of debris is formed (Hyodo et al., 2017b). When the midplane of the disk is initially not aligned with the equatorial plane of Mars, the dynamical flattening term $J_2$ of Mars' gravitational potential induces a precession of the disk particles symmetrically around the equatorial plane. Particle-particle inelastic collisions additionally damp their inclination, eventually forming an equatorial circular disk (Hyodo et al. 2017b).

SPH calculations modelling the Borealis-type impact indicate that the resulting disk would have a mass of several $10^{20}$ kg (Citron et al., 2015; Rosenblatt et al., 2016; Hyodo et al., 2017a), somewhat larger that the value estimated by Craddock (2011) and by Canup & Salmon (2018), about $10^{18} – 10^{19}$ kg. Crucially for the formation of Phobos and Deimos, the calculations also suggest that while most of the total ejected mass is confined below the Roche limit, roughly 1% lies beyond this, forming a tenuous outer disk of material that can extend past the synchronous limit (Rosenblatt et al., 2016; Canup & Salmon, 2018). The resolution of the SPH calculations however is not sufficient to provide a precise value for the outer disk mass, nor its density profile.

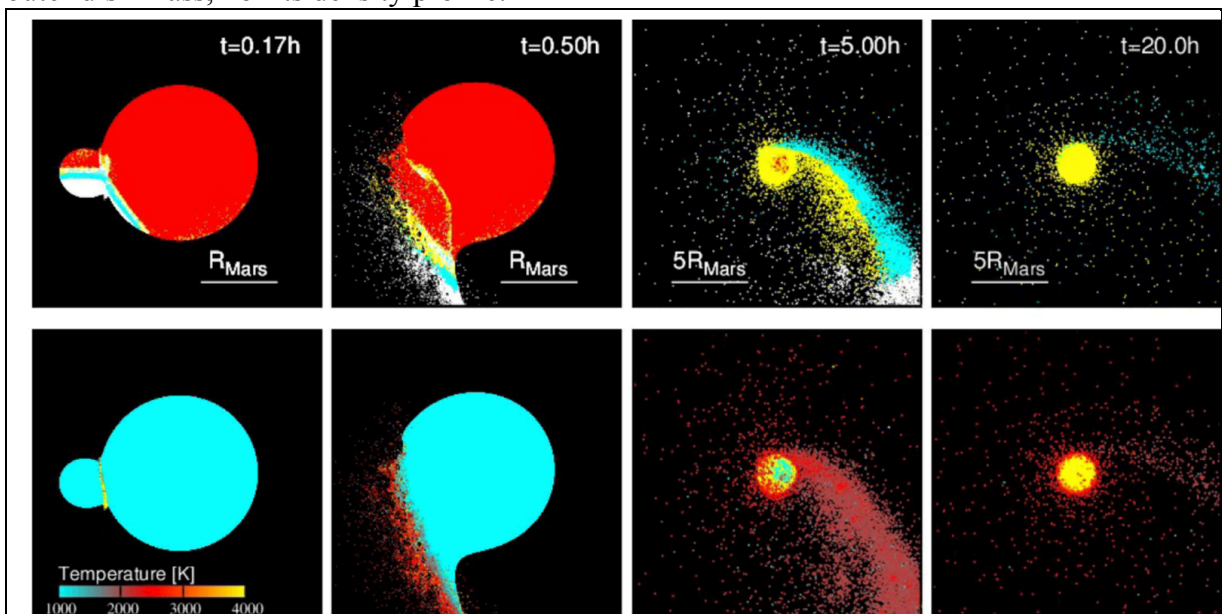

Figure 6: Snapshot of Smooth Particles Hydrodynamics (SPH) simulations of the Martian moons forming impact (adapted from Hyodo et al., 2017a). The orbital evolution of the orbit and temperature (color bar in Kelvin) of the blasted particles is shown on the top and bottom panels, respectively. The simulations run over 20 hours after the impact. The red, yellow, white and cyan dots of the top panel represent particles of Mars, falling on Mars, escaping from Mars, and forming a circum-Martian disk, respectively. Similar results have been obtained from other SPH simulations by Citron et al. (2015) and by Canup & Salmon

(2018). The latter authors however considered a lower impactor mass and thus a lower-energy impact.

Furthermore, some impact-debris could have escaped from Mars' gravity field and started to orbit around the sun. If this ejecta hit a primordial asteroid with a high impact velocity (> 5 km/s), the impact signatures (such as impact melt or/and 40Ar-39Ar resetting age) can be recorded (Hyodo & Genda 2018). Also, the Borealis-forming impact would excavate Martian mantle material (that is olivine-rich) and some of them are potentially implanted in the asteroid region as rare A-type asteroids (Polishook et al. 2017, Hyodo & Genda 2018).

*2-3  Chemistry of the debris cloud.*

As SPH computations show that the disk material is initially made of a mixture of gas (vapour) and melt (Hyodo et al., 2017a), Phobos and Deimos should be formed from the condensation of these two components. SPH calculations also suggest that the basic building blocks of the two moons are composed of roughly half-Martian and half-impactor material (Hyodo et al., 2017a), assuming a high-energy Borealis-forming impact (Marinova et al., 2008). A lower-energy impact could result in a higher proportion, up to 80%, of Martian material (Canup & Salmon, 2018).

As the disk cools, the gas condenses into small crystalline dust grains and the melt solidifies (Ronnet et al., 2016). Thermodynamic calculations coupled with dynamical modeling can be used to predict the composition of the building blocks that will accrete into moons (Visscher and Fegley, 2013). In particular, since the impactor may originate far from where Mars itself formed, the question arises as to whether differences in the impactor's chemistry (such as cometary or carbonaceous chondrite) leave a detectable trace in the composition of Phobos and Deimos (Craddock, 2011; Ronnet et al.,2016; Pignatale et al., 2018). This is an important question since, as mentioned in section: 'Are Phobos and Deimos small asteroids?', spectral observations of the surfaces of Phobos and Deimos are unable to clearly determine their composition. Reflectance spectra suggest some carbonaceous material (Fraeman et al., 2014), while emissivity spectra strongly point to the presence of silicates (Giuranna et al., 2011).

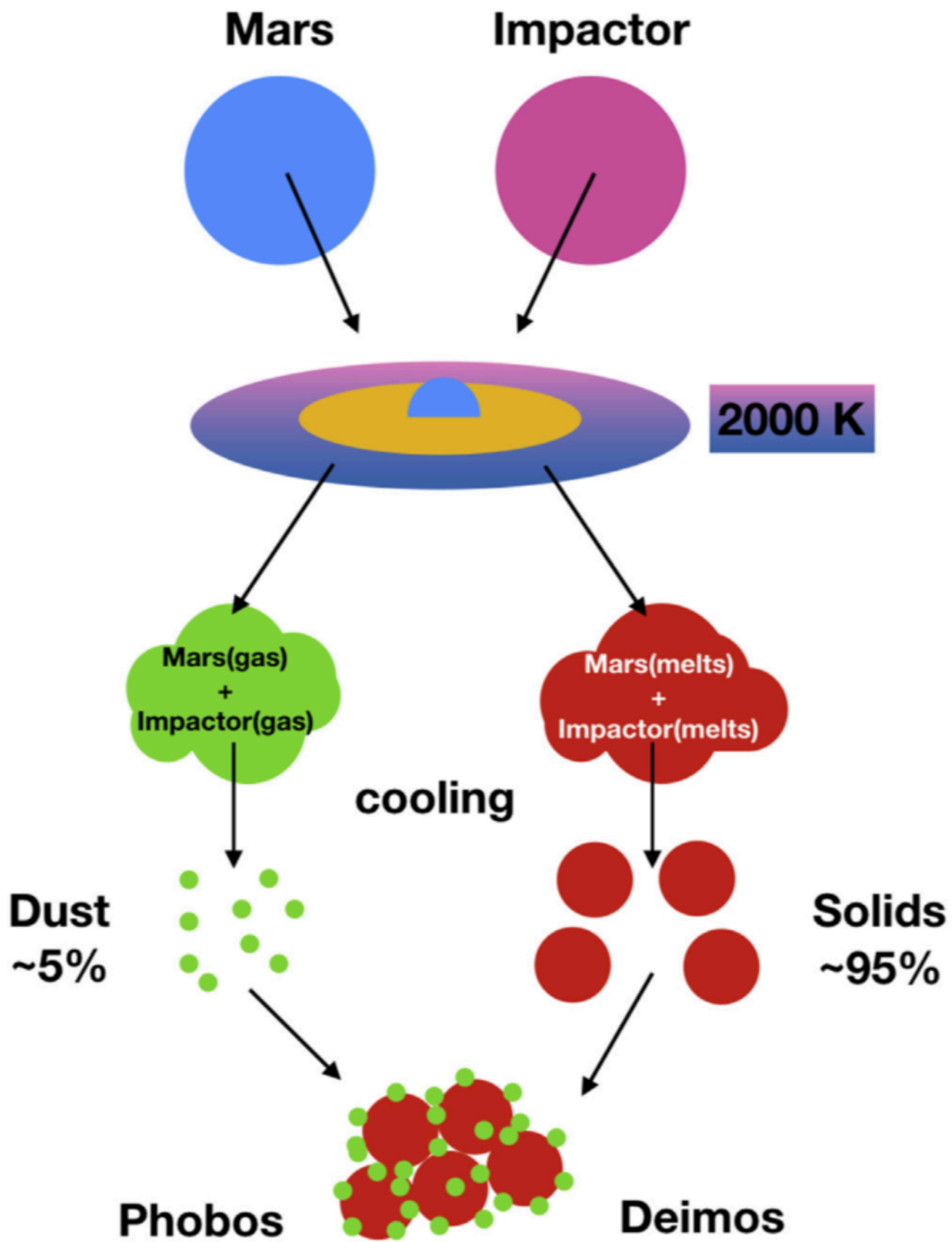

Figure 7: Schematic illustration of the chemical modeling of the giant impact scenario (from Pignatale et al., 2018). After the impact, some of the Martian material is ejected out at high temperature and vaporizes into gas, together with part of the impactor. The gas mixture then condenses into dust. On the other hand, the unvaporized material from Mars and the impactor forms a melt and then solidifies. Phobos and Deimos are the result of the accretion of these two components. The yellow region represents the part of the disk within the Roche limit (Hyodo et al. 2017a).

The thermodynamic calculations, using different types of impactor, predict a great diversity of final compositions for impactors carrying different amounts of C, H, O, Fe, Si; changing the proportion of these elements can significantly modify the resulting chemistry of the dust and melt (Pignatale et al., 2018). For example, a CV-chondrite type (anhydrous carbonaceous) type impactor would bring metallic iron, silica, iron sulfides and carbon; a cometary object would bring the largest carbon and water ice content; an enstatite chondrite would bring the largest quantities of sulfides.

The presence or absence of compounds such as metallic iron, iron oxides and iron-rich silicates, sulfides, carbon and water ice can therefore hint at the nature of the impactor.

Most of the impactors (carbonaceous-, enstatite-, and mars-type) produces iron-rich dust and some (carbonaceous-, cometary-type) also produce substantial quantities of carbon-rich dust, an opaque material, which thus lowers the albedo of the surface of the accreted bodies. Furthermore, as the grain size does not exceed 0.1~10 microns (Hyodo et al., 2017a), the reflectance of the surface is reduced (Ronnet et al., 2016). The solidified material also tends to lack a perfect crystalline structure, reducing the reflectance further. The small amount of dust (at most 5%) in the final material may nonetheless explain the low reflectance spectra of the surface of Phobos and Deimos (see section: 'Are Phobos and Deimos small asteroids?' and figure 3).

The possible imperfect crystallisation of melt could result in compositional variability for the building blocks of Phobos (Pignatale et al., 2018), which may reflect the so-called blue and red spectral units observed at the surface of Phobos (Murchie et al., 1991, see also figure 2). Indeed, such chemical variability seems necessary in order to explain the puzzling stratigraphic relationship between the blue and red units observed throughout the area of the large impact crater Stickney (Basilevsky et al., 2014).

Moreover, all types of impactor composition considered in Pignatale et al. (2018), except the CI-chondrite type (water-rich carbonaceous), produce a low amount of iron-rich silicates in the dust, which could explain the emissivity signature of Phobos' surface (Giuranna et al., 2011). In addition, the melt concentrates minerals of silicate-rich minerals that may explain the good match between Phobos's emissivity spectra and those of silicate material (Giuranna, 2011).

The material condensed in Mars orbit after a giant impact hence may not be incompatible with the spectral observations of the surface of the Martian moons, thus would not require any asteroidal material formed beyond Mars' orbit to account for these spectral observations. However, more detailed simulations of reflectance and emissivity spectra for the predicted condensed material are needed to assess the matching with Phobos' and Deimos' spectra. These studies are challenging since the additional space weathering effect is only well known for silicate material. The matching of simulated and observed spectra can therefore yield ambiguous results (Gaffey, 2010).

In their giant collision model, Canup & Salmon (2018) predict a disk debris temperature similar to the one in Hyodo et al. (2017a) but a different Mars-to-impactor ratio of the debris composition (80% Mars – 20% impactor). The final composition however should be quite similar to that predicted by Pignatale et al. (2018) for a Mars-like impactor (which results in a full Mars' composition). Nevertheless, a smaller amount of iron-rich silicate dust should be present, thus decreasing the darkening of the reflectance spectra that would hinder a comparison with observed reflectance spectra.

If the giant impact hypothesis is correct, depletion of volatile such as water vapour may occur since the impact is generally energetic (Hyodo et al. 2017a, Nakajima and Canup, 2017; Hyodo et al. 2018). Hyodo et al. (2018) consider two possible mechanisms for volatile depletion: hydrodynamic escape of vapour and blow-off of the volatile-rich condensates from the vapour by radiation pressure.

The vapour temperature just after the impact is $T_{VAP} \sim 2000$ K and the orbits of the debris are highly eccentric (Hyodo et al., 2017a), increasing their chance of escaping the system as the distance to Mars becomes larger. SPH calculations indicate that 10-40 % of the vapour satisfies the escape conditions during the first orbit from the impact point, depending on the impactor composition and vapour temperature between (1000-2000 K). Since the vapour contains more volatile elements than the melt, some fraction of the volatile may be lost from the original abundance by hydrodynamic escape (Hyodo et al., 2018).

Along the trajectory of the debris from impact point to apocenter (at the farthest distance from Mars), some fraction of the vapour may also condense and form small volatile-rich dust particles. The heat on Mars' surface generated by the impact (2000-6000 K) is strong enough to blow-off these small dust particles; "moderately" volatile elements, whose the condensation temperature is of 700-2000 K, and whose ratio of radiation pressure to gravitational forces is larger than 0.1, are most likely to be removed by radiation pressure (Hyodo et al. 2018). This loss of volatile elements has to be taken into account in any prediction of the final composition of the material condensed in Mars orbit from a giant collision.

*2-4  Evolution of the accretion disk*

SPH calculations of the initial impact and the nascent accretion disk are computationally very demanding, and 3D hydrodynamic simulations are therefore limited to 10-20 hours after the impact. This is however sufficient to study the orbital distribution of the blasted material around Mars that will then form the accretion disk (see section: 'Post-impact dynamical evolution of the debris cloud blasted in Mars' orbit: Formation of an accretion disk' and figure 6). Describing the long-term evolution of this multi-fluids disk (vapor mixed with solids) with a hydrodynamics code is however not feasible with the computing resources available today.

SPH calculations however provide material radial distributions whose density decreases with distance away from Mars. The calculations do not have enough resolution to provide fine details of the disk structure. Nevertheless, most of the mass is clearly concentrated below the Roche limit, and can form a dense inner disk, while the rest of the material, which extends slightly beyond the synchronous limit, forms a low-density outer disk (Rosenblatt et al., 2016; Hyodo et al., 2017; Canup and Salmon 2018).

The physics of the dense inner disk is similar to that of a viscous fluid (Salmon et al., 2010). The strong effective viscosity results from the formation of spiral self-gravitating wakes, owing to the large surface density of the inner disk. As a result, the inner disk spreads outwards from the planet. When material crosses the Roche limit, it can accrete into individual moons (low mass satellites, Charnoz et al., 2010). The orbital evolution of these moons is then driven by two opposite forces due to their gravitational interaction with the inner disk material (which transfers angular momentum to the moon orbits, thus repelling them) and with the planet (through tidal dissipation which reduces the angular momentum of the moon orbit, see section: 'Orbital evolution of the Martian satellites after the accretion period'). The former interaction pushes the orbit of each moonlet outwards while the second pulls it back towards the planet when the orbit is below the synchronous limit. Beyond this limit, the two forces act in the same direction and the moonlet migrates outwards definitively, but very slowly. This mechanism has successfully explained the formation of the small moons of Saturn from the planets's ring system, where the synchronous limit is in fact slightly below the Roche limit (Charnoz et al., 2010).

In the case of Mars, the synchronous limit (6 $R_M$) is far beyond the Roche limit (about 2.45 $R_M$). After the giant impact, the inner disk is massive, and disk-satellite interactions dominate planet-satellite tidal interactions. Thus, any moonlet forming at the Roche limit is initially pushed outward. However, as the disk's edge is located at the Roche limit, disk-satellite interactions cannot push a satellite beyond the synchronous limit; the maximal outward migration distance is defined by the 2:1 Lindblad resonance with the outer edge of the inner disk (i.e. the Roche limit), which is about 4.5 $R_M$ (Charnoz et al., 2010; Rosenblatt & Charnoz, 2012).

Over time, the dense inner disk empties, losing material either inwards to the surface of Mars or outwards across the Roche limit. As the disk density decreases, the Lindblad resonances become weaker (i.e. disk-satellites interactions decrease) and tidal dissipation eventually dominates, causing the orbits of any moonlet below the synchronous limit to decay. Eventually all these moonlets disappear below the Roche limit. The lifetime of the moonlet system depends on the mass of the inner disk; for a disk of mass $10^{19}$ kg, moonlets with the mass of Phobos of Deimos can form, but they completely disappear after 200 Ma (Rosenblatt & Charnoz, 2012), which is much lower than the presumed age of Phobos' surface, which is estimated as old as 4 billion of years (Schmedemann et al., 2014). A more massive disk would evolve even faster and would produce more massive moonlets, while a lighter disk would evolve more slowly but would produce moonlets less massive than Phobos and Deimos.

As the moons cross back below the Roche limit, they do not necessarily crash onto the surface of Mars. They can be disrupted by the tidal forces of Mars, breaking apart to form a new inner disk less massive than the initial disk (Black & Mittal, 2015). The whole process of moon formation can then restart from this new less massive disk. It has been suggested that Phobos today is the result of the latest iteration, 2.5 billion years ago, of such a cycle of disk formation and dispersal processes, initially triggered by a giant impact (Hesselbrock and Minton, 2017). However, a caveat of this scenario is the absence of a faint remnant ring around Mars.

These studies however do not take into account the low-density outer disk whose existence is suggested by SPH simulations. The question then is, could Phobos and Deimos have formed from the material in this outer disk?

*2-5   How to form two small outer satellites from a circum-Martian accretion disk: A dynamical solution.*

Since the density of the outer disk is low, it can be represented by a set of small bodies, or satellite embryos, rather than as a viscous fluid as was done for the dense inner disk. The evolution of these satellite embryos can be followed by numerically integrating the N-body equations of motion. A collision is commonly deemed to have occurred if two bodies approach each other to within their mutual Hill radius. The collision is treated as inelastic, resulting in accretion if the relative rebound velocity is smaller than the mutual escape velocity. One caveat is that disruption (i.e. the break-up of one or both colliding bodies) is neglected, based on the assumption that the resulting fragments remain close to each other and thus quickly recombine.

N-body simulations of the evolution of such an outer disk usually result in a stable configuration involving typically a dozen or so small moons whose orbits change little so that there are no more collisions (a dynamically 'frozen' system). The masses and orbits of these moons reflect the initial density profile of the disk due to the conservation of the center of mass (Rosenblatt et al. 2016). For instance, the moons tend to be homogeneously distributed if the accretion disk is uniform. In order to enhance accretion efficiency, and hence the possibility of forming Phobos and Deimos with their actual masses at the expected distances from Mars, it is necessary to somehow dynamically excite the outer disk.
Such an excitation is provided by the outward migration of the more massive transient inner moons that form at the Roche limit. The satellite embryos in the outer disk can be trapped in mean-motion resonances with the inner moons (when the ratio of their orbital period is a ratio of two integers), and follow concurrently their outward migration. The accumulation of embryos in a resonance favours accretion through collisions, while collisions also provide a mechanism of escaping the resonance. In approximately one third of the results reported by Rosenblatt et al (2016), two moons were formed by this mechanism, the more massive lying below the synchronous orbit and the less massive lying above it as it is for the Martian system (table 1 & 2).

Since more massive inner moons migrate more quickly, the rate of collisions and hence the possibility of accretion can increase, which can lead to cases where all the debris in the outer disk has been accreted onto one or more of the inner moons. This in turn suggests a limit for the mass of the inner disk and hence of the initial impactor (Canup and Salmon, 2018). These authors place this limit at around $2x10^{21}$ kg (impactor) and $2x10^{19}$ kg (disk) compared to the values of $2x10^{22}$ kg (impactor) and $5x10^{20}$ kg (disk) in Rosenblatt et al. (2016). Although the initial mass of the disk, and thus the energy of the impact in both studies differ significantly, the basic process of forming moons from re-accretion of debris after a giant collision with Mars is similar (Figure 8).

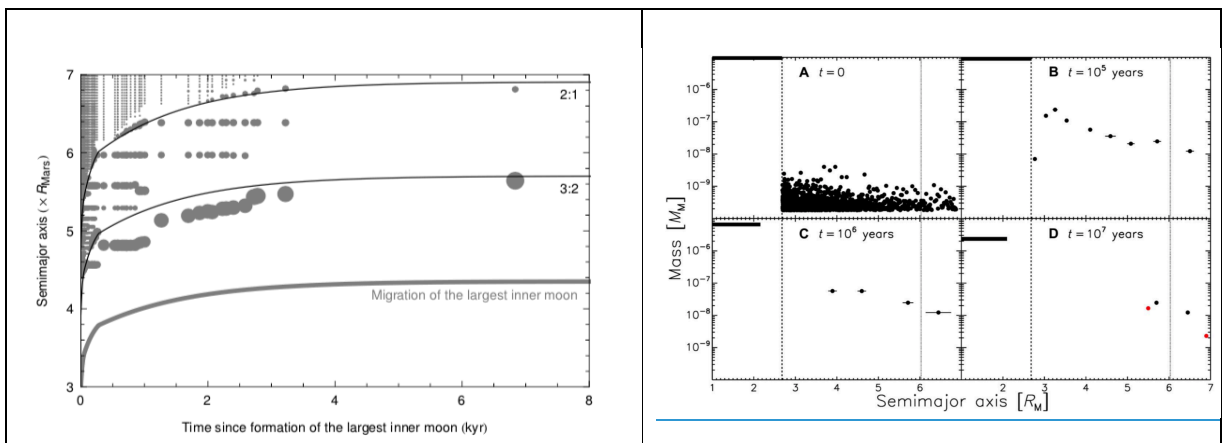

Figure 8: Accretion of satellite-embryos in the outer part of the accretion disk. (Left) from Rosenblatt et al. (2016): the accretion of embryos is facilitated by Mean Motion Resonances with the largest inner moon (1000 times the mass of Phobos) as it migrates outwards, requiring a high-energetic impact such as the canonical widely accepted

Borealis-forming impact. After about 8000 years only two small satellites survive on each side of the synchronous limit at 6 $R_M$. (Right) from Canup & Salmon (2018): the accretion of embryo follows same physical modeling as in Rosenblatt et al. (2016) but all inner moons and their resonance interactions are taken into account and the tidal dissipation of Mars is significantly higher. Similar results are obtained after 10 million years and a less energetic impact is required.

As well as increasing the semi-major axis of a trapped satellite embryo, mean-motion resonances can also increase its eccentricity. More massive inner moons migrate more quickly, leading to larger eccentricities for the satellite embryos. These can be reduced through collisions and accretion processes, but in numerical simulations (Rosenblatt et al., 2016) the final two outer moons often have eccentricities somewhat larger than those of Phobos and Deimos today. Tidal forces acting over billions of years can help damp these eccentricities, as discussed in the section: 'Orbital evolution of the Martian satellites after the accretion period'. Alternatively, if the mass of the initial impactor is smaller, the inner disk and hence the moons formed from it will be less massive. The outward migration would then be less rapid, the mean-motion resonances would have less effect which could result in more numerous small moons in the outer disk with smaller eccentric orbits. (Canup and Salmon, 2018).

*2-6   Orbital evolution of the Martian satellites after the accretion period*

If tidal evolution cannot allow capture scenarios (section 1.3), it also sets stringent constraints on accretion scenarios. Compared to other planets with moons in our solar system, for Mars the ratio of the centrifugal acceleration to gravitational acceleration is quite small, resulting in a relatively distant synchronous limit at about 6 $R_M$. In contrast to the Earth, this ratio may have barely changed throughout Mars' history (Dones and Tremaine, 1993). The Martian system therefore best illustrates the following rule (Murray and Dermott, 1999): moons above the synchronous limit recede away and siphon off angular momentum from the planet (Deimos behaves like most moons in the solar system, including Earth's Moon), while moons below the synchronous limit fall back and restore angular momentum to the planet (Phobos is the most notorious example, but Mars likely had many more moons in the past, see section: 'A giant collision scenario').

The precise orbital tidal evolution is controlled by gravitational torques from the tidal bulges raised on Mars by the moons, and from the tidal bulges raised on the moons by Mars. The evolution equations (Kaula, 1964) depend on a set of parameters $k_2/Q(\chi)$ describing tidal dissipation: $k_2$ is the degree-2 potential Love number and $Q(\omega)$ is the tidal quality factor at the principal tidal frequency $\chi$ (Efroimsky and Lainey, 2007). The Love number $k_2$ depends weakly on frequency; for Mars it can be estimated from tidal perturbations on the orbital motion of Martian spacecraft, which gives a $k_2$ value equal to 0.169 with an error of about 2% (Konopliv et al., 2016; Genova et al., 2016); for the moons, it has not been measured but models predict $k_2 \sim 10^{-4}$ for monoliths with a silicate composition (Lambeck 1979) and $k_2 \sim 10^{-3}$ for rubble piles also with a silicate composition; the porosity indeed decreases rigidity (Jaeger et al., 2007) and in turn increases $k_2$ (Rosenblatt et al.,2011; Le Maistre et al., 2013), which would be consistent with the high porosity inferred from the mass and volume of Phobos (Andert et al., 2010; Rosenblatt 2011). The quality factors $Q(\chi)$ depend strongly on the principal tidal frequency χ; for Mars it can be estimated at the current semi-diurnal frequency from the secular acceleration of the mean anomaly of Phobos' current orbit, which yields $Q = 82.8 \pm 0.2$ (Jacobson, 2010). The quality factors for Phobos and Deimos have not been measured but rubble piles are expected to be more dissipative than non-porous rocks, with $Q < 100$ (Goldreich and Sari, 2009).

In the course of tidal evolution, the orbit of the Martian moons shrinks or expands, according to whether it is within or beyond the synchronous limit, mainly due to dissipation within the planet, and circularises mainly due to dissipation within the moon (Goldreich, 1963); the inclination is barely affected (Mignard, 1981). As the orbit tidally evolves, it passes through several resonances, mainly between the mean motion and Mars's spin at 3.8 $R_M$ (a 2:1 resonance) and 2.9 $R_M$ (a 3:1 resonance), which will generally re-excite the eccentricity (Yoder, 1982). Other resonances, for example between the pericentre and Mars's mean motion, are harder to model (Yokoyama, 2002), given the chaotic variations of Mars' obliquity (Laskar and Robutel, 1993; Touma and Wisdom, 1993).

If Phobos and Deimos formed from a massive initial disc of debris, with a mass $\sim 10^{-3}$ $M_M$ (where $M_M$ is the mass of Mars), corresponding to an impactor of mass $\sim 10^{-1.5} M_M$ (Citron et al., 2015), mean motion resonances with large transient moons (see section: 'How to form two small outer satellites from a circum-Martian accretion disk: A dynamical solution') could facilitate the formation of exactly two moons with the correct semi-major axes, but with an eccentricity of around $10^{-2}$ for Deimos (Rosenblatt et al., 2016). Damping this eccentricity by tidal forces to that observed today would require $k_2/Q$ to be of the order of $10^{-4}$ for Deimos, at the upper limit of the expected range for rubble piles, and of the order of $10^{-6}$ for Phobos, at the lower limit of the expected range for monolithic rocks, thus implying a different structure or composition for the two moons. If Phobos and Deimos formed from a less massive initial disk of debris, with a mass or around $10^{-5}$ $M_M$, corresponding to an impactor of mass about $10^{-3}$ $M_M$, mean motion resonances would be largely ineffective, preventing undesirable excitation of eccentricities, but with the result that several moons could be left orbiting Mars (Canup and Salmon, 2018). For these to disappear, the value of $k_2/Q$ for Mars would have to be an order of magnitude larger than that presently observed, which makes the current orbital configuration with exactly two moons much less likely. The current resolution of SPH simulations is however too coarse to model the distribution of mass in the outer region of the initial disc of debris.

## 3- Conclusion

The formation of moons in the solar system is a long-standing topic of research (Peale & Canup, 2015). Each moon system has unique characteristics, making it difficult to envisage a common mechanism of formation. Recent spacecraft exploration of giant planets, in particular the Cassini mission around Saturn, have brought to light the role played by rings of debris and tidal forces on the formation and orbital evolution of small moons (Crida and Charnoz, 2012). Similar processes can explain the formation of Phobos and Deimos in a disk following a giant impact on Mars. A robust scenario must however be able to explain the formation of two small moons rather than a single massive one. The spin imparted to the planet by the impactor plays a major role, as this determines where the synchronous limit is, in particular if it is within or beyond the planet's Roche limit. A post-impact fast rotator will have a synchronous limit close to the planet, which favours the accretion of debris into a single body, as in the case of Earth's moon. A post-impact slow rotator such as Mars, with a spin of about 24 hours, corresponds to a relatively far synchronous limit; moons that form close to the Roche limit, by viscous spreading of the disk, migrate outwards through the interaction with the remnant disk, but under tidal forces they will eventually fall back towards Mars. Only small moons formed close to or beyond the synchronous limit are expected to remain in orbit for a long time. The current spin of Mars is an argument in favour of a relatively massive impactor, which could also be responsible for the formation of the Borealis basin (Hyodo et al., 2017b).

Numerical simulations for the complete scenario, from the initial impact to accretion and the long-term evolution of the two moons, yield better agreement with the present orbits than the previous widely-accepted capture hypothesis, and are still compatible with the observations concerning their composition. The amount of material blasted into orbit primarily depends on the mass of the impactor and on the angle of impact. This is crucial as it drives the evolution of the debris cloud into an accretion disk, as well as determining its eventual chemical composition. The various collision parameters are however difficult to constrain as they depend on the detailed dynamics of the early solar system (Hansen, 2018).

JAXA (Japan Aerospace eXploration Agency) plans a sample return mission from Phobos in the 2020s. The Martian Moons eXploration (MMX) probe will collect about 10 grams of material from Phobos' regolith. If the giant impact hypothesis is correct, MMX would collect not only material from the impactor but also from Mars, including ancient mantle material (Hyodo et al., 2017a). The detailed composition of these samples would help constrain the thermodynamical environment in which they formed and hence the parameters of the giant impact.

Further observations and analysis are thus required to test the giant impact hypothesis and future sample return missions such as MMX will give critical information about the composition (Murchie et al., 2014) and hence the origin of the Martian moons, Phobos and Deimos.

**Further Reading**

Scientific Papers:

The N-body problem:
    https://www.oxfordreference.com/view/10.1093/acref/9780191851193.001.0001/acref-9780191851193-e-2513

The Lindblad resonance:
    http://jeheo.weebly.com/uploads/5/3/8/0/53803559/lindblad_resonance_doyle_151119.pdf
    https://www.oxfordreference.com/view/10.1093/oi/authority.20110803100106597

The Solar system:
    https://books.google.fr/books?id=0bEMAwAAQBAJ&printsec=frontcover&hl=fr&source=gbs_ge_summary_r&cad=0#v=onepage&q&f=false

Movie & images:
    https://www.astro.oma.be/en/a-reappraisal-of-the-origin-of-mars-moons/
    https://www.youtube.com/watch?v=6Kn3wOWZg78
    https://www.esa.int/Our_Activities/Space_Science/Mars_Express/Mars_Express_heading_towards_daring_flyby_of_Phobos
    https://www.google.com/search?rls=en&q=Mars+Express+Phobos+flyby&tbm=isch&source=univ&client=safari&sa=X&ved=2ahUKEwimjKT-s8vjAhXB2-AKHTodBTcQsAR6BAgHEAE&biw=1324&bih=1038

Future exploration of Phobos & Deimos:
    http://mmx.isas.jaxa.jp/en/